\documentclass[12pt]{article}
\usepackage{amssymb,latexsym}

\begin{document}
\newcommand{\nn}{\nonumber}
\newcommand{\miso}{\frac{1}{2}}
\def\beq{\begin{equation}}
\def\eeq{\end{equation}}
\def\bea{\begin{eqnarray}}
\def\eea{\end{eqnarray}}
\def\mc{\mathcal}
\newcommand{\m}{\mathbf}
\newcommand{\fet}{\frac{1}{3}}
\newcommand{\fdt}{\frac{2}{3}}
\newcommand{\ftt}{\frac{4}{3}}
\def\w{\wedge}
\def\1{{G_2}/{SU(3)}}
\def\2{{Sp_4}/({SU(2)\times U(1))_{non-max}}}
\def\3{{SU(3)}/{U(1)\times U(1)}}
\def\olra{\overleftrightarrow}

\begin{titlepage}

\begin{flushright}
CERN-PH-TH/2010-035
\end{flushright}

\vspace{20pt}

\begin{center}

{ \Large{\bf Nearly-K\"{a}hler dimensional reduction of the heterotic string}}

\vspace{35pt}

{\bf Athanasios~Chatzistavrakidis}$^{1,2}$ {\bf and} {\bf G.~Zoupanos}$^{2,3}$ \vspace{20pt}

$^1$ {\it Institute of Nuclear Physics,
NCSR  Demokritos,\\
GR-15310 Athens, Greece}\\
 \vspace{5pt}

$^2${\it Physics Department, National Technical University of Athens, \\
GR-15780 Zografou Campus, Athens, Greece} \\
\vspace{5pt}

$^3${\it  Theory Group, Physics Department, CERN, Geneva, Switzerland}\\ \vspace{5pt}

\

Email: {\tt cthan@mail.ntua.gr, zoupanos@cern.ch}

\vspace{35pt}

{ABSTRACT}
\end{center}

 The effective action in four dimensions resulting from the ten-dimensional ${\cal N}=1$ heterotic supergravity coupled to ${\cal N}=1$ supersymmetric Yang-Mills upon dimensional reduction over nearly-K\"{a}hler manifolds is discussed. Nearly-K\"{a}hler manifolds are an interesting class of manifolds admitting an $SU(3)$-structure and in six dimensions all homogeneous nearly-K\"{a}hler manifolds are included in the class of the corresponding non-symmetric coset spaces plus a group manifold. Therefore it is natural to apply the Coset Space Dimensional Reduction scheme using these coset spaces as internal manifolds in order to determine the four-dimensional theory.

\vspace{20pt}

%\end{center}

\vspace{20pt}

\end{titlepage}

\newpage

\section{Introduction}

Superstring theories are the best candidates for a unified description of all fundamental interactions, including gravity. However, they are consistent only in ten dimensions and therefore it is crucial to find a suitable compactification which would lead to a realistic four-dimensional theory. This task has been pursued in many diverse ways over the last years.

The heterotic string theory \cite{Gross:1984dd} is one of the most popular versions of the superstring theory, since apart from the dimension of the higher-dimensional theory it also suggests its gauge group, which is $E_8\times E_8$ or $SO(32)$. The $E_8\times E_8$ gauge group is the most interesting case, since it can be broken down to phenomenologically interesting Grand Unified Theories (GUTs), where the standard model could in principle be accommodated. Moreover, the presence of chiral fermions in the higher-dimensional theory serves as an advantage in view of the possibility to obtain chiral fermions also in the four-dimensional theory.

The heterotic string theory is an ${\cal N}=1$ supersymmetric theory in ten dimensions and its dimensional reduction over tori leads in four dimensions to ${\cal N}=4$ supersymmetry, which is not a phenomenologically acceptable. The realization that Calabi-Yau (CY) threefolds serve as suitable compact internal spaces in order to maintain an ${\cal N}=1$ supersymmetry after dimensional reduction \cite{Candelas:1985en} led to pioneering studies in the dimensional reduction of superstring models \cite{Witten:1985xb,Derendinger:1985kk}. However, in CY compactifications the resulting low-energy field
theory in four dimensions contains a number of massless chiral
fields, known as moduli, which correspond to flat directions of the effective
potential and therefore their values are left undetermined.

The attempts to resolve the moduli stabilization problem led to the study of compactifications with fluxes (see \cite{Grana:2005jc} for a review). In this context the recent developments suggested the use of a wider class of internal spaces, called manifolds with $SU(3)$-structure, that contains CYs. Manifolds with $SU(3)$-structure have been exploited in supersymmetric type II compactifications \cite{Gurrieri:2002wz}-\cite{Cassani:2009ck} as well as in heterotic compactifications \cite{LopesCardoso:2002hd}-\cite{Benmachiche:2008ma}.

An interesting class of manifolds admitting an $SU(3)$-structure is that of nearly-K\"{a}hler manifolds. The homogeneous nearly-K\"{a}hler manifolds in six dimensions have been classified in \cite{Butruille:2006} and they are the three non-symmetric six-dimensional coset spaces and the group manifold $SU(2)\times SU(2)$. In the studies of heterotic
compactifications the use of non-symmetric coset
spaces was introduced in \cite{Govindarajan:1986kb}-\cite{Castellani:1986rg} and recently
developed further in \cite{LopesCardoso:2002hd,Manousselis:2005xa}.
Moreover, the effective theories resulting from dimensional reduction of the heterotic string over nearly-K\"{a}hler manifolds were studied in \cite{Chatzistavrakidis:2008ii,Chatzistavrakidis:2009mh}. Last but not least it is worth noting that the dimensional reduction of ten-dimensional ${\cal N}=1$ supersymmetric gauge theories over non-symmetric coset spaces led in four dimensions to softly broken ${\cal N}=1$ theories \cite{Manousselis:2000aj}.

\section{Nearly-K\"{a}hler manifolds}

Calabi-Yau manifolds are extensively used in string compactifications because they preserve ${\cal N}=1$ supersymmetry in four dimensions \cite{Candelas:1985en}. This feature stems from the existence of a nowhere-vanishing, globally defined spinor on these manifolds, which is covariantly constant with respect to the Levi-Civita connection. However, another possibility is to use a connection with torsion. The six-dimensional manifolds which admit a spinor which is covariantly constant with respect to the new connection are called manifolds with an $SU(3)$-structure. Their structure is completely specified by a real two-form $J$ and a complex three-form $\Omega$, which are both globally-defined and non-vanishing and they satisfy the compatibility conditions \beq J\w J\w J = \frac{3}{4}i\Omega\w\Omega^*, \qquad J\w\Omega=0. \eeq In addition, the exterior derivatives of these structure forms define the components ${\cal W}_i,i=1,\dots,5$ of the intrinsic torsion as
\bea\label{strfrms} dJ &=& \frac{3}{4}i({\cal
W}_1\Omega^*-{{\cal
W}}^*_1\Omega)+{\cal W}_4\w J+{\cal W}_3, \nn\\
    d\Omega &=& {\cal W}_1J\w J+{\cal W}_2\w J +{\cal W}_5^*\w
    \Omega. \eea
The intrinsic torsion classes ${\cal W}_i$ can be used to classify the several types of manifolds \cite{Grana:2005jc}. 

Nearly-K\"{a}hler manifolds are defined in the above context as the ones whose only non-vanishing torsion class is ${\cal W}_1$. The homogeneous nearly-K\"{a}hler manifolds in six dimensions have been classified in \cite{Butruille:2006} and they are the coset spaces
$\1,\2,\3$ and the group manifold $SU(2)\times SU(2)$.
The first three cases are well-known to be the only non-symmetric coset spaces $S/R$ in six dimensions which preserve the rank, namely $rankS=rankR$. They have been studied extensively in \cite{Kapetanakis:1992hf} in the reduction of ten-dimensional gauge theories to four dimensions. Therefore it is interesting to study the dimensional reduction of the heterotic supergravity-Yang-Mills theory over these spaces and determine the corresponding effective actions in four dimensions. In our examination we ignore the group manifold case since it cannot lead to chiral fermions in four dimensions.

In order to perform the dimensional reduction from ten to four dimensions, the forms on which the several fields will be expanded have to be specified. A natural basis of expansion forms consists of the $S$-invariant forms of the manifolds \cite{MuellerHoissen:1987cq}. Let us mention that all the non-symmetric coset spaces do not admit $S$-invariant one-forms. However, $S$-invariant two-forms, which we shall denote by $\omega_i$, exist in all cases and in particular there is one for $\1$, two for $\2$ and three for $\3$. Finally, all the three spaces admit two $S$-invariant three-forms, which we shall denote by $\rho_1$ and $\rho_2$. Forms of higher rank also exist but they are not important in our context. The above forms are related to the structure forms $J$ and $\Omega$ of the $SU(3)$-structure as
\beq J=R_i^2\omega_i, \hspace{5pt} \Omega = V(\rho_2+i\rho_1), \eeq where $R_i$ refers to the radii of the manifolds and $V$ to their volume. In particular, $\1$ admits only one radius $R_1$, $\2$ admits two radii $R_1,R_2$ and $\3$ admits three radii $R_1,R_2,R_3$. Therefore the volume is $R_1^3$, $R_1^2R_2$ and $R_1R_2R_3$ respectively in each case.

\section{Dimensional reduction}

The bosonic sector of the Lagrangian of the heterotic supergravity coupled to supersymmetric Yang-Mills \cite{Bergshoeff:1981um}, which is the low-energy
limit of the heterotic superstring theory, can be written as \beq\label{lagrangian}
\hat{e}^{-1}\mc{L}_{b}=
-\frac{1}{2\hat{\kappa}^2}\biggl(\hat{R}\hat{*}\mathbf{1}+ \frac{1}{2}
e^{- \hat{\phi}}\hat{H}_{(3)}\wedge \hat{*}\hat{H}_{(3)}+\frac{1}{2}
d\hat{\phi}\wedge
                \hat{*} d\hat{\phi}+\frac{\alpha'}{2}e^{-\frac{\hat{\phi}}{2}}Tr(\hat{F}_{(2)}\w\hat{\ast}\hat{F}_{(2)})\biggl).
                                               \eeq
The above Lagrangian, written in the Einstein frame, contains the ten-dimensional Einstein-Hilbert action, the kinetic term of the ten-dimensional dilaton $\hat\phi$, the kinetic term for the gauge fields $\hat A$
and the corresponding one for the three-form $\hat H$. The hats denote ten-dimensional fields, while $\hat{\kappa}$ is the
gravitational coupling constant in ten dimensions with
dimensions [length]$^4$; $\hat{e}$ is the
determinant of the metric, while $\hat\ast$ is the Hodge star
operator in ten dimensions.

In order to dimensionally reduce the above Lagrangian to four dimensions we perform the following Ans\"{a}tze for the fields appearing in (\ref{lagrangian}). The metric Ansatz is \begin{equation}
d\hat{s}^2 = e^{2\alpha\varphi(x)} \eta_{mn} e^{m} e^{n} +
e^{2\beta\varphi(x)}\gamma_{ab}(x)e^{a}e^{b},
\end{equation} where $e^{2\alpha\varphi(x)}\eta_{mn}$ is the four-dimensional metric and $e^{2\beta\varphi(x)}\gamma_{ab}(x)$ is the internal metric, while $e^m$ are the one-forms of the orthonormal basis in four dimensions and $e^a$ are the left-invariant one-forms on the coset space. The exponentials rescale the metric
components  in order to obtain an action
without any prefactor for the four-dimensional Einstein-Hilbert part. In order to achieve this we have to choose the values of $\alpha$ and $\beta$ to be $-\frac {\sqrt 3} {4}$ and $\frac{\sqrt 3}{12}$ respectively. Moreover, the
dilaton is trivially reduced by $\hat{\phi}(x,y) = \phi(x)$, since
it is already a scalar in ten dimensions. 

The three-form $\hat{H}$ is given in
general by \beq\label{hexp} \hat{H} =  \hat{d}\hat{B} -\frac{\alpha'}{2}
\hat{\omega}_{YM}, \eeq where $\hat{B}$ is the abelian
two-form potential, which we expand in the $S$-invariant forms of the internal space as
\begin{equation}\label{twoform}
\hat{B}= B(x) + b^i(x)\omega_i(y),
\end{equation} and $\hat{\omega}_{YM}$ is the Yang-Mills-Chern-Simons form. In (\ref{hexp}) a term involving the Lorentz-Chern-Simons form has to be included too. However, it is not needed in the minimal supergravity Lagrangian and therefore we shall not consider it here.

Finally, in order to reduce the gauge sector we employ the Coset Space Dimensional Reduction (CSDR) scheme \cite{Forgacs:1979zs,Kubyshin:1989vd,Kapetanakis:1992hf} (see also \cite{Lechtenfeld:2006wu} for an alternative approach based on equivariant reduction). The original CSDR of a multidimensional gauge field $\hat{A}$ on a
coset $S/R$ is described by a generalized invariance
condition
\begin{equation}\label{21}
{\cal L}_{X^{I}}\hat{A} = DW_{I}=dW_{I} + [\hat{A},W_{I}],
\end{equation}
where $W_{I}$ is a parameter of a gauge transformation associated
with the Killing vector $X_{I}$ of $S/R$ and ${\cal L}_{X^I}$ denotes the Lie derivative with respect to $X^I$. The Ansatz for the higher dimensional gauge field that solves the invariance condition (\ref{21}) is
\begin{equation}\label{apot}
\hat{A}^{\tilde{I}}=A^{\tilde{I}} + \phi^{\tilde{I}}_{A}e^{A},
\end{equation}
where $\tilde{I}$ is a gauge index and $A$ an $S$-index, which can
be split into indices $i,a$ running in the group $R$ and the coset respectively. The same Ansatz is used for the reduction of the $\hat \omega_{YM}$. It is important to mention that in the CSDR scheme the gauge group $G$ in ten dimensions is broken down to the centralizer $C_G(R)$ of the group $R$ in $G$ in four dimensions. In the present cases the initial $E_8\times E_8$ gauge group is broken down to $E_6\times E_8$ and therefore the resulting theories are ${\cal N}=1$ supersymmetric $E_6$ GUTs. Moreover, in the CSDR scheme the soft supersymmetry breaking sector of the four-dimensional ${\cal N}=1$ theories is obtained \cite{Manousselis:2000aj}.

The effective action in four dimensions is obtained by substituting the above expressions in the original ten-dimensional action. This procedure results in the Lagrangian
\beq\label{sugra4} {\cal
L}_b=-\frac{1}{2\kappa^2}R*1-\frac{1}{2}Re(f)F^I\w\ast
F^I+\frac{1}{2}Im(f)F^I\w {F}^I-\frac{1}{\kappa^2}G_{i\bar{j}}d\Phi^i\w\ast
d\bar{\Phi}^{\bar{j}}-V(\Phi,\overline{\Phi}). \eeq
In (\ref{sugra4}) $\kappa$ is the gravitational coupling in four dimensions, related to the ten-dimensional one by $\kappa^2=\frac{\hat{\kappa}^2}{V}$, $f$ is the gauge kinetic function and $G_{i\bar{j}}$ is the K\"{a}hler
metric. the potential has the form
\beq\label{spot}
V(\Phi,\bar{\Phi})=\frac{1}{\kappa^4}e^{\kappa^2K}\big(K^{i\bar{j}}
\frac{DW}{D\Phi^i}\frac{D\overline{W}}{D\bar{\Phi}^{\bar{j}}}-3\kappa^2W\overline{W}\big)+D-\mbox{terms},
\eeq where the derivatives involved are the K\"{a}hler covariant
derivatives, $W$ is the superpotential and by $\Phi^i$ we collectively denote the chiral supermultiplets. The superpotential $W$ can be determined by the Gukov-Vafa-Witten formula \cite{Gukov:1999ya},\cite{Gukov:1999gr}, which has the form
\beq\label{gukov} W=\frac{1}{4}\int_{S/R}\Omega\w(\hat{H}+idJ), \eeq while the K\"{a}hler potential can be determined as the sum of two terms $ K=K_S+K_{T}$,  which are given by the expressions \beq\label{kaehler} K_S = -ln(S+{S}^*),\hspace{5pt}
                    K_T = -ln\biggl(\frac 16\int_{S/R} J\w J\w J\biggl), \eeq where $S$ is the dilaton superfield. A third contribution to the K\"{a}hler potential due to the complex structure moduli is not present here since the $SU(3)$-structures we consider are real.

Applying (\ref{gukov}) to the cases of $\1,\2$ and $\3$ respectively, the corresponding superpotentials take the form
\bea W_1&=&{3}T_1-\sqrt{2}\alpha'd_{ijk}B^iB^jB^k, \\
 W_2&=&2T_1+T_2-\sqrt{2}\alpha'd_{ijk}B^iB^j\Gamma^k,\\
W_3&=&T_1+T_2+T_3-\sqrt{2}\alpha'd_{ijk}A^iB^j\Gamma^k,\eea where $T_i$ are the superfields of the geometric moduli, $A^i,B^i,\Gamma^i$ are the vector superfields and $d_{ijk}$ is the symmetric tensor of $E_6$. Also, applying (\ref{kaehler}) the corresponding K\"{a}hler potentials are
\bea
K_1=&-&ln(S+S^*)-3ln(T_1+T_1^*-2\alpha'B_iB^i),\\
K_2=&-&ln(S+S^*)-2ln(T_1+T_1^*-2\alpha'B_iB^i)-ln(T_2+T_2^*-2\alpha'\Gamma_i\Gamma^i),\nn \\\\
K_3=&-&ln(S+S^*)-ln(T_1+T_1^*-2\alpha'A_iA^i)-ln(T_2+T_2^*-2\alpha'B_iB^i)\nn\\
&-& ln(T_3+T_3^*-2\alpha'\Gamma_i\Gamma^i).\nn\\ \eea Finally, in all three cases the gauge kinetic function turns out to be $f(S)=S$.

\section{Conclusions}

In this contribution we have presented the main steps of the derivation of the four-dimensional effective action which results from the heterotic supergravity coupled to supersymmetric Yang-Mills theory from ten dimensions to four, using homogeneous six-dimensional nearly-K\"{a}hler manifolds as
internal spaces. Since the homogeneous nearly-K\"{a}hler manifolds in six dimensions are the three corresponding non-symmetric coset spaces plus a group manifold (which was not studied here because it cannot lead to chiral theories), we employed the Coset Space Dimensional Reduction scheme to reduce the gauge sector of the theory. Moreover, the other parts of the ten-dimensional theory were consistently reduced using appropriate Ans\"{a}tze which amount to expanding the fields in $S$-invariant forms of the coset spaces $S/R$. The resulting theories ${\cal N}=1$ supersymmetric $E_6$ GUTs and they also contain terms which could be interpreted as soft scalar masses and trilinear
soft terms in four-dimensions, in case the minimization of the full potential would lead to
Minkowski vacuum. Their superpotential is determined in a straightforward way via the heterotic Gukov-Vafa-Witten formula and their K\"{a}hler potential is determined using results of the special K\"{a}hler geometry. 

It would be interesting to study further the possibility to determine a Minkowski vacuum with stabilized moduli in this context. Several directions exist in order to explore this possibility, such as gaugino condensation (see e.g. \cite{Nilles:2004zg} and references therein) and the KKLT scenario \cite{Kachru:2003aw}. Moreover, the inclusion of more general fluxes combined with the ones already considered could serve as another possibility to determine such vacua \cite{Kaloper:1999yr,Derendinger:2004jn}.    

\paragraph{Acknowledgements:} This work is supported by the NTUA programmes for basic research PEVE 2008 and 2009, the European Union's RTN programme under contract MRTN-CT-2006-035505 and the European Union's ITN programme "UNILHC" PITN-GA-2009-237920.

\end{document}